\documentclass[letterpaper, 10pt, conference]{ieeeconf}
\IEEEoverridecommandlockouts

\usepackage{cite}
\usepackage{amsmath,amssymb,amsfonts}
\usepackage{algorithmic}
\usepackage{graphicx}
\usepackage{textcomp}
\usepackage{xcolor}
\usepackage{booktabs}
\usepackage{tikz}
\usetikzlibrary{shapes.geometric, arrows.meta, positioning, fit, calc}
\usepackage{url}
\let\labelindent\relax
\usepackage{enumitem}
\setlist{noitemsep,topsep=2pt}

\definecolor{cblue}{RGB}{41,98,163}
\definecolor{cgreen}{RGB}{46,139,87}
\definecolor{corange}{RGB}{204,102,0}
\definecolor{cred}{RGB}{178,34,34}

\def\BibTeX{{\rm B\kern-.05em{\sc i\kern-.025em b}\kern-.08em
    T\kern-.1667em\lower.7ex\hbox{E}\kern-.125emX}}

\begin{document}

\title{Consent Chain Degradation in Embodied Multi-Agent Systems: Bridging the Gap Between AI Agent Governance and Robot Ethics}

\author{Mehmet Haklidir$^{1}$%
\thanks{$^{1}$Mehmet Haklidir is with the Artificial Intelligence Institute, T\"{U}B\.{I}TAK B\.{I}LGEM, Kocaeli, Turkey
        {\tt\small mehmet.haklidir@tubitak.gov.tr}}%
\thanks{ORCID: 0000-0003-4985-1116}%
}

\maketitle

\begin{abstract}
Robotic systems are moving from isolated platforms to interconnected multi-agent ecosystems that operate in human environments. This shift raises a governance problem that existing frameworks do not address: how does consent propagate, degrade, and break down across chains of delegation between embodied autonomous agents? The AI ethics community has begun to study consent for digital software agents, and the HRI community has examined consent in dyadic human-robot encounters. Neither body of work covers what happens when physical robots delegate tasks to other robots in ways that affect humans. This paper introduces \textit{consent chain degradation} (CCD), a conceptual framework for analyzing how the specificity, validity, and scope of human consent erodes as authority passes through multi-robot delegation chains. We propose a three-layer governance architecture, the \textit{Consent Runtime Verification Framework for Embodied Agents} (CoRVE), which integrates consent scope modeling, delegation chain tracking, and physical irreversibility assessment. Three scenarios in healthcare, domestic, and industrial robotics show how CCD arises in practice, including a worked numerical example. A regulatory gap analysis covering the EU AI Act, the GDPR, the Machinery Regulation, and the Revised Product Liability Directive shows that all four instruments leave core CCD dimensions unaddressed.
\end{abstract}

\section{Introduction}

Robot deployment in human environments is growing. The International Federation of Robotics reported 4.3 million operational industrial robots worldwide in 2023, a 10\% year-over-year increase~\cite{ifr2024}. Social, surgical, and autonomous service robots are entering healthcare facilities, homes, and public spaces. The shift from single-robot deployments to interconnected multi-robot systems creates a new governance problem: when multiple robots coordinate on tasks that involve humans, who holds the authority to consent on behalf of the affected individuals?

Consent is central to medical ethics~\cite{beauchamp2019}, data protection law (e.g., the EU GDPR~\cite{gdpr2016}), and social contract theory. In the principlist framework of Beauchamp and Childress, consent is a direct expression of the principle of \textit{respect for autonomy}: a person's right to make informed decisions about actions that affect them. Arnold and Scheutz~\cite{arnold2019} established consent as a distinct research area in human-robot interaction (HRI), showing that it covers not only explicit permission but also implicit acquiescence in normatively neutral scenarios. Lintvedt~\cite{lintvedt2024} showed that conflating ``informed consent'' with data privacy consent obscures the distinct challenges posed by robots' physical embodiment.

The existing literature, however, treats consent almost exclusively in \textit{dyadic} human-robot interactions. LLM-based multi-agent systems for robotic coordination~\cite{mandi2025}, authenticated delegation frameworks for AI agents~\cite{authdel2025}, and multi-robot teaming architectures~\cite{scheutz2024} now create configurations where a human's original consent must pass through multiple autonomous agents before becoming a physical action. This paper addresses that gap with three contributions:

\begin{enumerate}
    \item We introduce \textbf{consent chain degradation} (CCD), a conceptual framework with formal notation for how consent erodes across multi-agent delegation chains in embodied systems (Section~\ref{sec:ccd}).
    \item We propose \textbf{CoRVE}, a three-layer architecture for runtime consent verification in embodied multi-agent systems (Section~\ref{sec:corve}).
    \item We analyze three scenarios that expose CCD, including a worked numerical example, and evaluate CoRVE against the current EU regulatory framework (Sections~\ref{sec:scenarios}--\ref{sec:regulatory}).
\end{enumerate}

\noindent We situate this work at a robotics venue because the physical irreversibility of robotic action (property P1) is the central differentiator between CCD and software-agent consent problems; this distinction disappears in venues that treat all AI agents alike regardless of embodiment.

\section{Background and Related Work}

\subsection{Consent in Human-Robot Interaction}

Arnold and Scheutz~\cite{arnold2019} mapped consent across social and legal doctrine, identifying empirical and technical questions for major HRI application domains. Their work showed that consent in HRI involves touch, proxemics, gaze, and moral norms, dimensions absent from standard informed consent models.

Lintvedt~\cite{lintvedt2024} showed that HRI research conflates informational and physical privacy dimensions of consent. In dementia care, meaningful consent is hard to obtain~\cite{ostlund2023, pyer2024}, and deception, autonomy loss, and emotional dependency raise separate ethical concerns~\cite{boada2021}. An ethical framework for human-robot collaboration in Industry 5.0 manufacturing identified the ``illusion of informed consent'' as a problem: workers may formally consent to collaborative robotic environments without understanding the implications~\cite{industry5ethics2024}. The HRI Value Compass~\cite{hricompass2025} turned ethical values into a practical design instrument, though its developers noted a Western-centric bias in HRI ethics research. Coeckelbergh's Ubuntu robot framework~\cite{coeckelbergh2024} addresses this bias by proposing relational rather than individualistic governance models.

\subsection{The Responsibility Gap and Its Consent Parallel}

Matthias~\cite{matthias2004} introduced the ``responsibility gap'': learning automata create situations where neither manufacturer nor operator can fully predict machine behavior. Sparrow~\cite{sparrow2007} applied the argument to autonomous weapons. Santoni de Sio and Mecacci~\cite{santoni2021} refined it into four distinct gaps (culpability, moral accountability, public accountability, and active responsibility). Danaher~\cite{danaher2022} argued that responsibility gaps can sometimes be desirable because they relieve the burden of tragic moral choices.

We argue that an analogous \textit{consent gap} arises in multi-agent robotic systems. The responsibility gap describes situations where no identifiable agent bears moral responsibility for outcomes. The consent gap describes situations where no identifiable consent, human or delegated, authorizes the specific physical actions a robot performs. In Beauchamp and Childress's terms~\cite{beauchamp2019}, CCD is a violation of \textit{respect for autonomy}: when consent degrades through a delegation chain, the affected person's capacity for self-determination over their body and environment is undermined without their knowledge.

\subsection{Autonomy, Delegation, and Trust}

Beer et al.~\cite{beer2014} proposed a 10-point taxonomy of robot autonomy levels (LORA). A recent framework from the Knight First Amendment Institute~\cite{loa2025} defined five levels of AI agent autonomy, extending to multi-agent systems where agents can themselves be users of other agents. Cantucci et al.~\cite{cantucci2025} showed empirically (N=373) that robot competence is the primary driver of trust-based task delegation. This raises a concern: humans may delegate authority based on perceived competence rather than on the actual scope of their consent.

Falcone and colleagues' cognitive architecture for collaborative autonomy~\cite{falcone2022} lets robots adjust their autonomy level based on attributed mental states. This introduces the \textit{consent inference problem}: if a robot infers a user's goals and acts without being asked, has implicit consent been given? Hou, Cheon, and Jung~\cite{hou2024} showed that power dynamics, though pervasive in HRI, remain largely unexplored, with direct implications for how consent is negotiated or coerced.

\subsection{Multi-Agent Systems and Robotic Coordination}

LLM-based multi-agent systems now coordinate robotic perception, planning, and execution~\cite{mandi2025}. Authenticated delegation frameworks~\cite{authdel2025} propose verifiable authority chains using OAuth 2.0 extensions but target digital agents only. The distinct challenge of \textit{embodied} delegation, where delegated actions have physical and sometimes irreversible consequences, remains open.

\section{Consent Chain Degradation}
\label{sec:ccd}

The ``illusion of informed consent'' identified in manufacturing HRI~\cite{industry5ethics2024} shows that consent can be structurally inadequate even in single-robot settings. CCD generalizes this problem to multi-agent chains, where each delegation step introduces further opportunities for consent to lose its meaning.

\subsection{Definition}

\textbf{Consent chain degradation (CCD)} is the progressive loss of specificity, validity, and contextual appropriateness of human consent as it propagates through a chain of delegations between autonomous agents, especially when those agents are embodied and capable of physical action.

Formally, let $C_0$ be the original consent granted by human $H$ to robot $R_1$ for action scope $S_0$. When $R_1$ delegates a sub-task to $R_2$, the effective consent becomes $C_1 \subseteq S_0$, mediated by $R_1$'s interpretation. At each delegation step $i$:

\begin{equation}
    C_i = f_i(C_{i-1}, \delta_i, \epsilon_i)
\end{equation}

\noindent where $f_i$ is the delegation function of agent $i$, $\delta_i$ is the \textit{delegation drift} (divergence between the delegating agent's intent and the receiving agent's interpretation), and $\epsilon_i$ is the \textit{environmental perturbation} (contextual changes since consent was given). We present Eq.~(1) as structured notation for reasoning about CCD rather than as a directly computable function; concrete instantiations of $f_i$ will depend on the specific multi-robot architecture and consent representation used.

\subsection{Properties of CCD}

Four properties distinguish CCD from analogous problems in digital agent systems:

\textbf{P1: Physical Irreversibility.} Many robotic actions cannot be undone. A surgical incision is permanent; administered medication cannot be recovered. Consent violations in embodied systems therefore carry higher stakes than in software agent systems.

\textbf{P2: Temporal Decay.} Consent is time-bound. A patient who consented at $t_0$ may withdraw consent at $t_1$ because circumstances changed. Multi-agent chains widen the gap between consent granting and action execution.

\textbf{P3: Contextual Opacity.} As delegation depth grows, downstream agents lose access to the original consent context: the conditions, limits, and preferences that accompanied $C_0$. This parallels the opacity problem in the responsibility gap literature~\cite{matthias2004}, here applied to consent rather than causation.

\textbf{P4: Scope Ambiguity.} Natural language consent (e.g., ``the robot can help me in the kitchen'') is inherently ambiguous. That ambiguity compounds through delegation chains. Each agent's interpretation may be individually reasonable yet collectively exceed the human's intended authorization.

\subsection{CCD Severity Metric}

We propose a composite indicator $\Gamma$ for assessing CCD risk at runtime:

\begin{equation}
    \Gamma = \alpha \cdot \text{IR}(a) + \beta \cdot \hat{\Delta t} + \gamma \cdot \hat{D} + \lambda \cdot A(S)
\end{equation}

\noindent where $\text{IR}(a) \in [0,1]$ is the irreversibility index of physical action $a$, $\hat{\Delta t} = (t_n - t_0)/T_{\max} \in [0,1]$ is the temporal gap normalized by a domain-specific maximum validity period $T_{\max}$, $\hat{D} = D/D_{\max} \in [0,1]$ is the delegation depth normalized by a domain-specific maximum chain length $D_{\max}$, $A(S) \in [0,1]$ is the scope ambiguity, and $\alpha + \beta + \gamma + \lambda = 1$ are domain-specific weights. All terms are normalized to $[0,1]$, so $\Gamma \in [0,1]$. When $\Gamma$ exceeds threshold $\Gamma^*$, the system triggers re-consent or human-in-the-loop verification. This indicator is designed for structuring risk assessment; calibrating the weights and threshold for specific domains requires empirical work (see Section~\ref{sec:limits}).

\section{CoRVE: Consent Runtime Verification for Embodied Agents}
\label{sec:corve}

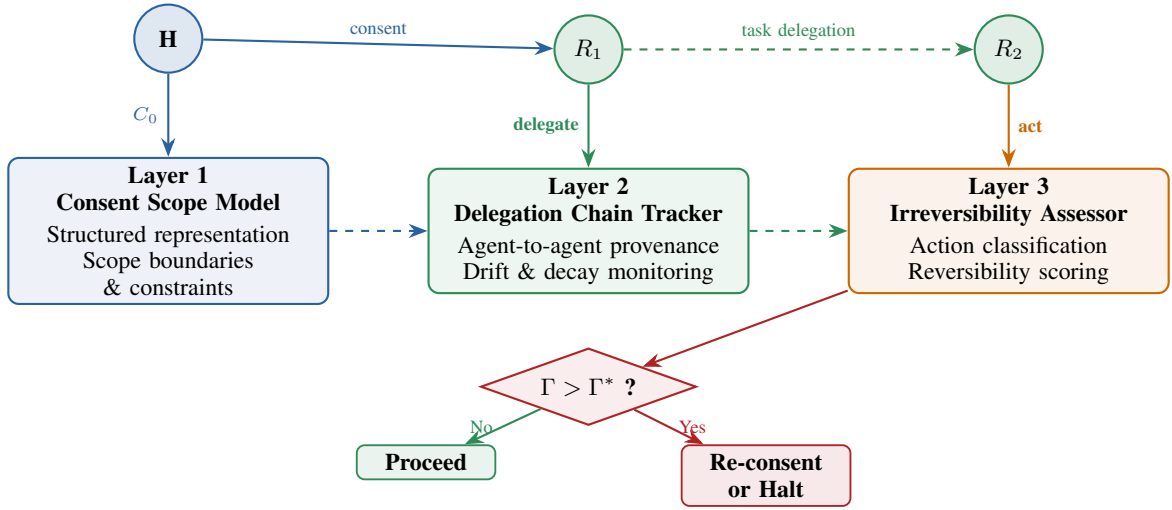
\begin{figure*}[!t]
\centering
\begin{tikzpicture}[
    node distance=0.7cm and 1.0cm,
    box/.style={rectangle, rounded corners=4pt, draw=#1, fill=#1!8, line width=0.8pt, text width=4.0cm, minimum height=1.5cm, align=center, font=\small},
    agent/.style={circle, draw=#1, fill=#1!15, minimum size=0.9cm, line width=0.8pt, font=\small\bfseries},
    arrow/.style={-{Stealth[length=2.5mm]}, thick, #1},
    lbl/.style={font=\scriptsize\bfseries, #1}
]
\node[box=cblue] (scope) {\textbf{Layer 1}\\\textbf{Consent Scope Model}\\[2pt] Structured representation\\Scope boundaries \& constraints};
\node[box=cgreen, right=1.3cm of scope] (chain) {\textbf{Layer 2}\\\textbf{Delegation Chain Tracker}\\[2pt] Agent-to-agent provenance\\Drift \& decay monitoring};
\node[box=corange, right=1.3cm of chain] (phys) {\textbf{Layer 3}\\\textbf{Irreversibility Assessor}\\[2pt] Action classification\\Reversibility scoring};
\node[agent=cblue, above=1.1cm of scope] (human) {H};
\node[agent=cgreen, above=1.1cm of chain] (r1) {$R_1$};
\node[agent=cgreen, above=1.1cm of phys] (r2) {$R_2$};
\node[diamond, draw=cred, fill=cred!8, minimum size=0.8cm, below=0.7cm of chain, font=\small\bfseries, aspect=2.8, line width=0.8pt] (decide) {$\Gamma > \Gamma^*$ ?};
\node[rectangle, rounded corners=3pt, draw=cgreen, fill=cgreen!10, below left=0.5cm and 0.5cm of decide, font=\small, text width=1.6cm, align=center, line width=0.7pt] (proceed) {\textbf{Proceed}};
\node[rectangle, rounded corners=3pt, draw=cred, fill=cred!10, below right=0.5cm and 0.5cm of decide, font=\small, text width=2.0cm, align=center, line width=0.7pt] (reconsent) {\textbf{Re-consent}\\\textbf{or Halt}};
\draw[arrow=cblue] (human) -- node[left, lbl=cblue]{$C_0$} (scope);
\draw[arrow=cgreen] (r1) -- node[left, lbl=cgreen]{delegate} (chain);
\draw[arrow=corange] (r2) -- node[right, lbl=corange]{act} (phys);
\draw[arrow=cblue, dashed] (scope) -- (chain);
\draw[arrow=cgreen, dashed] (chain) -- (phys);
\draw[arrow=cred] (phys) -- (decide);
\draw[arrow=cgreen] (decide) -- node[left, font=\scriptsize]{No} (proceed);
\draw[arrow=cred] (decide) -- node[right, font=\scriptsize]{Yes} (reconsent);
\draw[arrow=cgreen, dashed] (r1) -- node[above, font=\scriptsize]{task delegation} (r2);
\draw[arrow=cblue] (human) -- node[above, font=\scriptsize]{consent} (r1);
\end{tikzpicture}
\caption{CoRVE architecture. Human $H$ grants consent $C_0$ to $R_1$, which delegates to $R_2$. Three layers continuously assess consent validity. When CCD severity $\Gamma$ exceeds threshold $\Gamma^*$, the system triggers re-consent or halts execution.}
\label{fig:corve}
\end{figure*}

We propose the \textit{Consent Runtime Verification Framework for Embodied Agents} (CoRVE), a three-layer architecture that detects and mitigates CCD in real time. Fig.~\ref{fig:corve} gives an overview.

\subsection{Layer 1: Consent Scope Model (CSM)}

The CSM converts natural language or interaction-based consent into a structured representation using a tuple formalism:

\begin{equation}
    C = \langle H, \mathcal{A}, \mathcal{S}, T, \mathcal{E}, P \rangle
\end{equation}

\noindent where $H$ is the consenting human, $\mathcal{A}$ is the set of authorized action types, $\mathcal{S}$ is the spatial scope (rooms, zones, body regions), $T = [t_{\text{start}}, t_{\text{end}}]$ is the temporal validity window, $\mathcal{E}$ is the set of exclusion conditions (actions explicitly not consented to), and $P \in \{0,1\}$ is the delegation permission flag (whether consent may be further delegated).

$P$ defaults to 0 (no delegation) unless explicitly set otherwise. This follows a privacy-by-design principle analogous to GDPR's data minimization (Art.~5(1)(c))~\cite{gdpr2016}. The formalism yields machine-readable consent boundaries that can be checked at each delegation step.

\subsection{Layer 2: Delegation Chain Tracker (DCT)}

The DCT maintains a provenance graph of every delegation event. Each edge records the delegating and receiving agent identifiers, the consent scope subset being delegated, the timestamp and environmental context, and the computed delegation drift $\delta_i$.

The DCT monitors for \textit{scope creep}, that is, situations where a sequence of individually minor delegations produces actions outside the original consent scope:

\begin{equation}
    \text{ScopeCreep}(n) = 1 - \frac{|C_n \cap S_0|}{|C_n|}
\end{equation}

\noindent where $C_n$ is the effective consent at depth $n$ and $S_0$ is the original scope, both represented as sets of authorized action-space-time triples drawn from $\mathcal{A} \times \mathcal{S} \times T$. Values near 1 signal large divergence from the original consent.

\subsection{Layer 3: Physical Irreversibility Assessor (PIA)}

The PIA classifies each physical action on an irreversibility spectrum:

\begin{itemize}
    \item \textbf{Tier 1 -- Reversible} ($\text{IR} < 0.3$): Actions fully undoable without residual effects (e.g., moving a cup, opening a door).
    \item \textbf{Tier 2 -- Partially Reversible} ($0.3 \leq \text{IR} < 0.7$): Actions with mitigable residual effects (e.g., administering non-critical medication, rearranging furniture).
    \item \textbf{Tier 3 -- Irreversible} ($\text{IR} \geq 0.7$): Actions that cannot be undone (e.g., surgical incision, discarding items, physical contact with a vulnerable person).
\end{itemize}

For Tier 3 actions, CoRVE requires direct human re-consent \textit{regardless} of the $\Gamma$ score. This hard constraint overrides indicator-based decisions.

\subsection{Integration Summary}

Table~\ref{tab:layers} maps each CoRVE layer to its expected inputs and outputs in a robotic system.

\begin{table}[b]
\renewcommand{\arraystretch}{1.15}
\centering
\caption{CoRVE Layers: Inputs, Outputs, and Robot Subsystems (illustrative; mappings are architecture-dependent)}
\label{tab:layers}
\small
\begin{tabular}{@{}p{1.1cm}p{2.2cm}p{1.8cm}p{1.6cm}@{}}
\toprule
\textbf{Layer} & \textbf{Input} & \textbf{Output} & \textbf{Subsystem} \\
\midrule
L1: CSM & NL consent, gesture, context & Consent tuple $C$ & NLU, dialog manager \\
L2: DCT & Agent IDs, scope subsets, timestamps & Provenance graph, scope creep score & Task planner, middleware \\
L3: PIA & Action primitive, object class & IR score, tier label & Motion planner, world model \\
\midrule
\multicolumn{2}{@{}l}{Combined $\Gamma$ evaluator} & Proceed / re-consent / halt & Behavior executive \\
\bottomrule
\end{tabular}
\end{table}

\section{Scenario Analysis}
\label{sec:scenarios}

Three scenarios illustrate how CCD arises and how CoRVE responds. Table~\ref{tab:scenarios} provides a cross-scenario summary of which CCD properties and CoRVE layers are activated.

\begin{table}[tb]
\renewcommand{\arraystretch}{1.15}
\centering
\caption{CCD Properties and CoRVE Layer Activation per Scenario}
\label{tab:scenarios}
\small
\begin{tabular}{@{}lccc@{}}
\toprule
 & \textbf{S1: Health} & \textbf{S2: Home} & \textbf{S3: Industry} \\
\midrule
P1: Irreversibility & \checkmark & \checkmark & -- \\
P2: Temporal decay & \checkmark & -- & -- \\
P3: Context opacity & \checkmark & -- & \checkmark \\
P4: Scope ambiguity & -- & \checkmark & \checkmark \\
\midrule
L1: Scope Model & \checkmark & \checkmark & -- \\
L2: Chain Tracker & \checkmark & -- & \checkmark \\
L3: Irreversibility & \checkmark & \checkmark & \checkmark \\
\midrule
Outcome & re-consent & halt & notify \\
\bottomrule
\end{tabular}
\end{table}

\subsection{Scenario 1: Multi-Robot Healthcare (Worked Example)}

\textit{Context:} Patient P consents to robot nurse $R_N$ assisting with daily care. $R_N$ delegates medication delivery to pharmacy robot $R_P$, which delegates physical administration to bedside robot $R_B$.

\textit{CCD Manifestation:} P's original consent to $R_N$ covered ``daily care assistance.'' Through the chain $R_N \rightarrow R_P \rightarrow R_B$, the scope narrowed to medication administration while irreversibility (Tier 2--3) increased. Hours may separate consent from action. $R_B$ has no access to P's original preferences, emotional state, or the conversational context in which consent was given.

\textit{Worked $\Gamma$ computation.} Suppose a healthcare deployment sets $\alpha{=}0.4$, $\beta{=}0.2$, $\gamma{=}0.2$, $\lambda{=}0.2$ (weighting irreversibility highest), with $T_{\max}{=}8$h and $D_{\max}{=}5$. For $R_B$ administering oral medication: $\text{IR}{=}0.6$ (Tier 2), $\Delta t{=}3$h so $\hat{\Delta t}{=}0.375$, $D{=}3$ so $\hat{D}{=}0.6$, $A(S){=}0.3$ (``daily care'' is moderately ambiguous). Then $\Gamma = 0.4(0.6) + 0.2(0.375) + 0.2(0.6) + 0.2(0.3) = 0.495$. With $\Gamma^*{=}0.45$, this triggers re-consent. Had the action been moving a pillow ($\text{IR}{=}0.1$, $\hat{D}{=}0.4$, same $\hat{\Delta t}$, $A(S){=}0.1$), $\Gamma{=}0.215$, and the action would proceed.

\textit{CoRVE Response:} Layer 1 flags that medication administration is not in $\mathcal{A}$. Layer 2 detects a delegation depth of 3 with scope drift. Layer 3 classifies the action as Tier 2+. Combined, $\Gamma$ exceeds $\Gamma^*$, and the system requests re-consent from P.

\subsection{Scenario 2: Domestic Robot Ecosystem}

\textit{Context:} Resident R tells home hub robot $R_H$ to ``keep the house tidy.'' $R_H$ delegates kitchen cleaning to $R_K$ and instructs $R_K$ to coordinate with delivery robot $R_D$ to dispose of expired items.

\textit{CCD Manifestation:} R's consent has high scope ambiguity ($A(S) \approx 0.8$). The delegation chain introduces property disposal, an irreversible action that most humans would not consider implied by ``tidying.'' The chain $R_H \rightarrow R_K \rightarrow R_D$ shows compounding scope ambiguity (P4).

\textit{CoRVE Response:} Layer 1 detects high $A(S)$. Layer 3 classifies item disposal as Tier 3. Because the action is Tier 3, CoRVE halts disposal regardless of $\Gamma$ and asks R for explicit consent, listing the specific items flagged.

\subsection{Scenario 3: Collaborative Industrial Robotics}

\textit{Context:} Worker W consents to collaborative robot $R_C$ sharing their workspace. $R_C$ communicates with logistics robot $R_L$ to reroute material flow, changing W's workspace layout without notice.

\textit{CCD Manifestation:} W consented to $R_C$'s co-presence, not to workspace reconfiguration by $R_L$. The delegation from $R_C$ to $R_L$ happened through machine-to-machine communication invisible to W, creating contextual opacity (P3). This scenario is directly relevant to the collaborative risk mapping requirements in Annex III of the Machinery Regulation~\cite{machreg2023}.

\textit{CoRVE Response:} Layer 2 identifies $R_L$ as outside the original consent scope's agent set. Layer 3 classifies workspace reconfiguration as Tier 2 (partially reversible but safety-relevant). CoRVE notifies W and requires acknowledgment before $R_L$ proceeds.

\section{Regulatory Gap Analysis}
\label{sec:regulatory}

\begin{table}[tb]
\renewcommand{\arraystretch}{1.2}
\centering
\caption{Regulatory Coverage of CCD Dimensions}
\label{tab:regulatory}
\small
\begin{tabular}{@{}p{2.1cm}cccc@{}}
\toprule
\textbf{CCD Dim.} & \textbf{AI Act} & \textbf{GDPR} & \textbf{Mach.R.} & \textbf{PLD} \\
\midrule
Consent scope & $\circ$ & $\circ$ & -- & -- \\
Chain tracking & -- & -- & -- & $\circ$ \\
Phys.\ irrevers. & -- & -- & $\bullet$ & $\bullet$ \\
Runtime verif. & $\circ$ & -- & -- & -- \\
Multi-agent & -- & -- & -- & -- \\
Temporal decay & -- & $\circ$ & -- & -- \\
Re-consent & -- & $\circ$ & -- & -- \\
\bottomrule
\multicolumn{5}{@{}p{8.5cm}@{}}{\scriptsize $\bullet$ = regulation explicitly creates an obligation addressing this dimension; $\circ$ = regulation contains provisions relevant to but not specifically targeting this dimension; -- = no provision bears on this dimension.}
\end{tabular}
\end{table}

Table~\ref{tab:regulatory} maps CCD dimensions against four EU regulatory instruments.

\textbf{GDPR}~\cite{gdpr2016}. The GDPR provides the most developed legal model of consent. Art.~4(11) requires consent to be freely given, specific, informed, and unambiguous. Art.~7 requires that consent be withdrawable at any time. Art.~22 restricts automated decision-making affecting individuals. These provisions partially cover consent scope (through the specificity requirement), temporal decay (through withdrawability), and re-consent (through the right to withdraw). The GDPR's consent model, however, was designed for data processing, not for physical actions by embodied agents. It does not address delegation of consent between autonomous systems, and its notion of a ``data controller'' does not straightforwardly map onto multi-robot delegation chains where no single agent controls the full action sequence. When robots collect biometric or health data during interaction (e.g., a care robot monitoring vital signs), delegation to a second robot may trigger GDPR Art.~9 requirements for sensitive data processing. Art.~9(2) provides legal bases beyond consent (e.g., vital interests under Art.~9(2)(c)) that may apply in healthcare CCD scenarios, complicating a purely consent-centric governance approach.

\textbf{EU AI Act}~\cite{euaiact2024}. Art.~14 requires human oversight for high-risk AI systems, and Art.~13 requires transparency sufficient for users to interpret system output. Art.~9 mandates risk management systems. These partly cover consent scope modeling (through transparency) and runtime verification (through post-market monitoring). The Act does not address multi-agent consent chains or how consent propagates when an AI system classified as high-risk delegates to another system.

\textbf{Machinery Regulation}~\cite{machreg2023}. This regulation, applicable from 20 January 2027, addresses physical safety through risk assessment and introduces autonomy thresholds for self-evolving machinery. Its Annex III includes requirements for collaborative risk mapping in shared human-robot workspaces. These are \textit{safety} obligations, however, not \textit{consent} obligations: the regulation requires manufacturers to assess hazards from human-robot co-presence, but it does not require robots to verify that the humans in the workspace have consented to the specific actions being performed. It takes a meta-regulatory approach, mandating risk assessment processes rather than specific consent behaviors~\cite{mahler2024}.

\textbf{Revised PLD}~\cite{pld2024}. Directive (EU) 2024/2853 extends product liability to software and AI, partly covering delegation chains through expanded supply chain accountability. It operates post-hoc: it assigns liability after harm but does not provide preventive consent governance. The concept of ``defect'' under the PLD was not designed for distributed causation across delegation chains where cumulative drift, rather than a single malfunction, produces the harm.

None of these four instruments addresses multi-agent consent propagation as a distinct governance problem. This is the gap CoRVE targets. Non-EU jurisdictions also lack coverage: the US relies on sector-specific regulation without a comprehensive AI consent framework, and China's PIPL addresses data consent but not physical robotic delegation.

\section{Discussion}

\subsection{Relationship to Prior Work}

This work extends two prior threads. The ``consent vacuum''~\cite{haklidir2026vacuum} proposed a taxonomy of consent failure modes in digital multi-agent settings. The present paper extends that taxonomy to embodied systems, adding physical irreversibility (P1) and temporal decay (P2) as failure modes specific to robotics. Consent as a runtime safety constraint~\cite{haklidir2026runtime}, originally a token-level verification mechanism for LLM agents, is here adapted into the three-layer CoRVE architecture for multimodal robotic consent.

The responsibility gap literature~\cite{matthias2004, sparrow2007, santoni2021} provides the philosophical base, but the distinction matters: the responsibility gap concerns \textit{ex post} attribution of blame, while the consent gap concerns \textit{ex ante} authorization of action. CoRVE aims to prevent consent violations, not to assign responsibility after harm occurs. CoRVE's re-consent mechanism can also be read as an operationalization of ``meaningful human control''~\cite{santoni2018mhc}, ensuring that humans retain decision authority even when actions are initiated by delegated agents.

\subsection{Limitations and Open Questions}
\label{sec:limits}

\textbf{Status of the formalism.} The notation in Sections III--IV is conceptual scaffolding, not a directly executable specification. Implementing CoRVE (e.g., as ROS~2 middleware) requires concrete choices about consent representation, scope comparison, $\Gamma$ calibration, multimodal sensor fusion, NLU for scope extraction, and real-time computation within physical control latency bounds. A concrete prototype could be realized as a set of ROS~2 lifecycle nodes, with the CSM exposing consent tuples as JSON-LD or RDF resources, the DCT maintaining a provenance graph in a distributed key-value store, and the PIA acting as a behavior tree precondition node that gates action execution. The $\Gamma$ evaluation would then run as a pre-execution check within the behavior executive, with a target latency in the tens of milliseconds per delegation event to remain compatible with typical robotic control loops. Initial weight calibration could draw on Delphi-style elicitation with regulators, clinicians, and operators, refined post-deployment through Bayesian updates on logged re-consent outcomes. A staged validation roadmap would move from controlled lab studies through high-fidelity simulation to IRB-approved user studies in the target domains.

\textbf{Cultural adaptation.} The consent model assumes individual authorization in the Western liberal tradition. Collectivist cultures may need communal models~\cite{coeckelbergh2024} that the single-human tuple cannot capture. Extending to group consent ($H \rightarrow \{H_1, \ldots, H_k\}$ with a consensus function) is needed for global deployment.

\textbf{Consent fatigue.} Too-frequent re-consent requests cause habituation and rubber-stamping. Calibrating $\Gamma^*$ to balance safety against usability requires empirical work. Adaptive thresholds that learn from individual behavior may help.

\textbf{Inferred consent and coercion.} CoRVE assumes that $C_0$ was explicitly and freely granted. It cannot detect if the user was coerced due to power asymmetries (e.g., eldercare, workplace settings). Separately, robots with Theory of Mind capabilities may \textit{infer} consent from behavior or contextual cues~\cite{falcone2022}, as emergency medicine relies on presumed consent~\cite{beauchamp2019}. If $C_0$ was inferred rather than given, the delegation chain rests on an unverified foundation, a failure mode we call consent \textit{fabrication}.

\textbf{Consent withdrawal and latency.} CoRVE does not yet model active consent \textit{withdrawal}. Propagating withdrawal or halt signals through an active delegation chain requires real-time guarantees across distributed middleware (e.g., ROS~2 DDS) that remain an open engineering challenge.

\section{Conclusion}

This paper introduced consent chain degradation (CCD) for analyzing how human consent erodes across multi-agent delegation chains in embodied robotic systems. Four properties distinguish CCD from digital-agent consent problems: physical irreversibility, temporal decay, contextual opacity, and scope ambiguity. The severity indicator $\Gamma$ supports runtime risk assessment, and the CoRVE architecture provides real-time detection and mitigation. Scenario analysis and regulatory gap mapping showed that CCD constitutes a governance gap not covered by current EU regulations.

Future work should address: (1)~CoRVE implementation as ROS~2 middleware~\cite{macenski2022}, (2)~culturally adaptive consent models supporting communal authorization, (3)~integration into standards (IEEE P7001, whose decision-chain transparency aligns with DCT provenance tracking; P7017; ISO 13482), and (4)~empirical studies of consent fatigue and threshold calibration.

\section*{Acknowledgments}
Claude (Anthropic) was used only as a writing and formatting assistant to help improve language and clarity. All ideas, methodology, analysis, interpretations, figures, tables, and conclusions are the sole work of the author, who also reviewed and approved the final manuscript.


\begin{thebibliography}{34}

\bibitem{ifr2024}
International Federation of Robotics, \textit{World Robotics 2024: Industrial Robots}.\hskip 1em plus 0.5em minus 0.4em Frankfurt: IFR, 2024.

\bibitem{beauchamp2019}
T.~L.~Beauchamp and J.~F.~Childress, \textit{Principles of Biomedical Ethics}, 8th~ed.\hskip 1em plus 0.5em minus 0.4em Oxford University Press, 2019.

\bibitem{gdpr2016}
European Union, ``Regulation (EU) 2016/679 on the protection of natural persons with regard to the processing of personal data (GDPR),'' \textit{Official J. EU}, L~119, pp.~1--88, 2016.

\bibitem{arnold2019}
T.~Arnold and M.~Scheutz, ``When exceptions are the norm: Exploring the role of consent in HRI,'' \textit{ACM Trans. Human-Robot Interact.}, vol.~8, no.~3, Art.~14, 2019.

\bibitem{lintvedt2024}
M.~N.~Lintvedt, ``A critical analysis of consent in human--robot interaction,'' in \textit{The Cambridge Handbook of the Law, Policy, and Regulation for Human--Robot Interaction}, W.~Barfield, Y.-H.~Weng, and U.~Pagallo, Eds.\hskip 1em plus 0.5em minus 0.4em Cambridge, U.K.: Cambridge Univ. Press, 2024, ch.~17.

\bibitem{mandi2025}
Z.~Mandi, S.~Jain, and S.~Song, ``RoCo: Dialectic multi-robot collaboration with large language models,'' in \textit{Proc. IEEE Int. Conf. Robot. Autom. (ICRA)}, 2024, pp.~286--299.

\bibitem{authdel2025}
S.~Shavit \textit{et~al.}, ``Authenticated delegation and authorized AI agents,'' \textit{arXiv preprint arXiv:2501.09674}, 2025.

\bibitem{scheutz2024}
M.~Scheutz, B.~Oosterveld, J.~Peterson, E.~Wyss, and E.~Krause, ``A multi-robot architectural framework for effective robot teammates in mixed-initiative teams,'' in \textit{Proc. Int. Symp. Technol. Advances in Human-Robot Interact. (TAHRI)}, 2024, pp.~74--82.

\bibitem{ostlund2023}
L.~\"{O}stlund, M.~Ernsth~Bravell, and L.~Johansson, ``Working in a gray area: Healthcare staff experiences of receiving consent when caring for persons with dementia,'' \textit{Dementia}, vol.~22, no.~1, pp.~144--160, 2023.

\bibitem{pyer2024}
M.~Pyer and A.~Ward, ``Developing a dementia friendly approach to consent in dementia research,'' \textit{Aging \& Mental Health}, vol.~28, no.~2, pp.~294--301, 2024.

\bibitem{boada2021}
J.~P.~Boada, B.~R.~Maestre, and C.~T.~Gen\'{i}s, ``The ethical issues of social assistive robotics: A critical literature review,'' \textit{Technol. in Soc.}, vol.~67, p.~101726, 2021.

\bibitem{industry5ethics2024}
C.~Dignum \textit{et~al.}, ``An ethical framework for human-robot collaboration for the future people-centric manufacturing,'' \textit{Technol. in Soc.}, vol.~98, p.~102288, 2024.

\bibitem{hricompass2025}
L.~Sion \textit{et~al.}, ``Concerns and values in human-robot interactions: A focus on social robotics,'' \textit{arXiv preprint arXiv:2501.05628}, 2025.

\bibitem{coeckelbergh2024}
M.~Coeckelbergh, ``The Ubuntu robot: Towards a relational conceptual framework for intercultural robotics,'' in \textit{The Cambridge Handbook of the Law, Policy, and Regulation for Human--Robot Interaction}, W.~Barfield, Y.-H.~Weng, and U.~Pagallo, Eds.\hskip 1em plus 0.5em minus 0.4em Cambridge, U.K.: Cambridge Univ. Press, 2024, pp.~408--420.

\bibitem{matthias2004}
A.~Matthias, ``The responsibility gap: Ascribing responsibility for the actions of learning automata,'' \textit{Ethics Inf. Technol.}, vol.~6, no.~3, pp.~175--183, 2004.

\bibitem{sparrow2007}
R.~Sparrow, ``Killer robots,'' \textit{J. Appl. Philosophy}, vol.~24, no.~1, pp.~62--77, 2007.

\bibitem{santoni2021}
F.~Santoni~de~Sio and G.~Mecacci, ``Four responsibility gaps with artificial intelligence,'' \textit{Philosophy \& Technol.}, vol.~34, pp.~1057--1084, 2021.

\bibitem{danaher2022}
J.~Danaher, ``Tragic choices and the virtue of techno-responsibility gaps,'' \textit{Philosophy \& Technol.}, vol.~35, Art.~26, 2022.

\bibitem{beer2014}
J.~M.~Beer, A.~D.~Fisk, and W.~A.~Rogers, ``Toward a framework for levels of robot autonomy in human-robot interaction,'' \textit{J. Human-Robot Interact.}, vol.~3, no.~2, pp.~74--99, 2014.

\bibitem{loa2025}
K.~Feng \textit{et~al.}, ``Levels of autonomy for AI agents,'' Knight First Amendment Institute, Columbia University, 2025.

\bibitem{cantucci2025}
F.~Cantucci, M.~Marini, and R.~Falcone, ``The role of robot competence, autonomy, and personality on trust formation in human-robot interaction,'' \textit{ACM Trans. Human-Robot Interact.}, vol.~15, no.~3, Art. no.~55, pp.~1--27, 2026.

\bibitem{falcone2022}
R.~Falcone, A.~Sapienza, F.~Cantucci, and M.~Marini, ``Collaborative autonomy: Human-robot interaction to the test of intelligent help,'' \textit{Electronics}, vol.~11, no.~19, p.~3065, 2022.

\bibitem{hou2024}
Y.~T.-Y.~Hou, E.~Cheon, and M.~F.~Jung, ``Power in human-robot interaction,'' in \textit{Proc. ACM/IEEE Int. Conf. Human-Robot Interact. (HRI)}, 2024, pp.~269--282.

\bibitem{euaiact2024}
European Union, ``Regulation (EU) 2024/1689 laying down harmonised rules on artificial intelligence (AI Act),'' \textit{Official J. EU}, L series, 2024.

\bibitem{machreg2023}
European Union, ``Regulation (EU) 2023/1230 on machinery products,'' \textit{Official J. EU}, L~165, pp.~1--102, 2023.

\bibitem{mahler2024}
T.~Mahler, ``Smart robotics in the EU legal framework: The role of the Machinery Regulation,'' \textit{Oslo Law Review}, vol.~11, no.~1, pp.~1--18, 2024.

\bibitem{pld2024}
European Union, ``Directive (EU) 2024/2853 on liability for defective products,'' \textit{Official J. EU}, L series, 2024.

\bibitem{haklidir2026vacuum}
M.~Haklidir, ``The consent vacuum: Governing agent-to-agent interaction,'' submitted for publication, 2026.

\bibitem{haklidir2026runtime}
M.~Haklidir, ``Consent as a runtime safety constraint for LLM agents,'' submitted for publication, 2026.

\bibitem{torras2024}
C.~Torras, ``Ethics of social robotics: Individual and societal concerns and opportunities,'' \textit{Annu. Rev. Control, Robot., Auton. Syst.}, vol.~7, pp.~1--18, 2024.

\bibitem{santoni2018mhc}
F.~Santoni~de~Sio and J.~van~den~Hoven, ``Meaningful human control over autonomous systems: A philosophical account,'' \textit{Frontiers in Robot. and AI}, vol.~5, Art.~15, 2018.

\bibitem{macenski2022}
S.~Macenski, T.~Foote, B.~Gerkey, C.~Lalancette, and W.~Woodall, ``Robot Operating System 2: Design, architecture, and uses in the wild,'' \textit{Science Robotics}, vol.~7, no.~66, eabm6074, 2022.

\end{thebibliography}
\end{document}